\documentclass[12pt]{article}

\usepackage{pifont}
\usepackage{mathrsfs}
\usepackage{amsfonts}
\usepackage{color}
\usepackage{footmisc}
\usepackage[numbers,sort&compress]{natbib}
\usepackage[bookmarksopen=true, bookmarksopenlevel=\maxdimen,
bookmarksnumbered=true,colorlinks,linkcolor=blue,
citecolor=blue,urlcolor=blue]{hyperref}
\newcommand{\omits}[1]{}

\definecolor{darkgreen}{rgb}{0.0,0.6,0.0}
\definecolor{lightgrey}{rgb}{0.7,0.7,0.7}

\begin{document}
\begin{center}
{\bf \LARGE A Kaluza--Klein-like model of \\ \bigskip
the gauge theory of gravity\\ \bigskip and its cosmological meaning}

\bigskip

\bigskip

{\Large Jia-An Lu$^{a}$\footnote{Email: ljagdgz@163.com}}

\bigskip

$^a$ School of Physics and Engineering, Sun Yat-sen
University,\\ Guangzhou 510275, China

\begin{abstract}
A new Kaluza--Klein-like (KK-like) model of the de Sitter gauge theory of gravity
is constructed from the geometry related to the gauge-invariant expressions of
the gravitational fields. The model reduces to general relativity with a cosmological
constant $\Lambda$ when the spin tensor is equal to zero. Moreover, it is shown that
among the $R-2\Lambda+\beta S^{abc}S_{abc}$ models, where $\beta$ is a dimensionless
parameter and $S_{abc}$ denotes torsion, the KK-like model is the only one that may
avert the big-bang singularity of the Robertson--Walker universe filled with a
spin fluid.
\end{abstract}
\end{center}



\section{Introduction}

It has been pointed out that there are three kinds of special relativity
(SR) \cite{BdS,dSSR,SR-wick,PoR-SR}
and gravity should be based on the localization of a SR with full symmetry
\cite{PoI,Guo07,SR-Gravity}. It is a motivation for the study of the
Poincar\'e, de Sitter (dS) or Anti-de Sitter (AdS) gauge theory of gravity
\cite{Hehl76,Guo76,Stelle,Grignani,Lu13}, where the Riemann--Cartan (RC)
geometry with nontrivial metric and torsion is introduced to realize the
corresponding gauge symmetry.

The Kaluza--Klein-like (KK-like) models \cite{Mansouri,Guo79,Lu13} of the gauge theory
of gravity are some models where the Lagrangian densities are constructed from the scalar
curvatures of some fiber bundles, just like the KK theory \cite{Cho,Percacci}.
The scalar curvature of the fiber bundle is given by the RC geometries of the spacetime
and the typical fiber. In these models, one can see what effects the geometry of the typical fiber,
which is determined by the gravitational gauge group, would have on the gravitational dynamics.

In the previous KK-like model of the dS gauge theory of gravity, in order to make the
no-gravity spacetime be a vacuum solution, the Lagrangian density of a global section in the bundle
is introduced to give a suitable constant term \cite{Lu13}. However, the physical meaning of the
global section is unclear. In fact, the global section may just correspond to the local
inertial coordinates. In this paper, we introduce a coupling constant between the spacetime
and the typical fiber. By adjusting this coupling constant, we can make the no-gravity spacetime
be a vacuum solution, without introducing the Lagrangian density of the global section. The resulting
model is equivalent to general relativity (GR) with a cosmological constant when the spin tensor
is equal to zero. Furthermore, we show that among the $R-2\Lambda+\beta S^{abc}S_{abc}$ models,
where $\beta$ is a dimensionless parameter and $S_{abc}$ denotes torsion, the new KK-like model
is the only one that may avoid the initial singularity of the Robertson--Walker (RW) universe
filled with a spin fluid.

The paper is organized as follows. In section 2, the dS gauge
theory of gravity with the gauge-invariant expressions for the spacetime
metric, torsion and curvature is briefly introduced. In section 3, we construct
a new KK-like model of the dS gauge theory of gravity. In section 4, it is shown that
among a group of models, the new KK-like model is the only one that may avert the big-bang
singularity of the RW universe filled with a spin fluid. We give some remarks in section 5 and
end with three appendixes on the geometry related to the gauge-invariant expressions of
the metric, torsion and curvature.

\section{de Sitter gauge theory of gravity}

Let ${\cal M}$ be the spacetime manifold, with the metric $g_{ab}$,
and the metric-compatible derivative operator $\nabla_a$,
where $a,b$ are abstract indices \cite{Wald,Liang}.
There exist tensor fields $S^{c}{}_{ab}$ and $R^c{}_{dab}$, such that
for any local function $f$ and one-form field $\omega_{a}$ on ${\cal M}$,
\begin{equation}\label{torsion}
(\nabla_{a}\nabla_{b}-\nabla_{b}\nabla_{a})f=
-S^{c}{}_{ab}\nabla_{c}f,
\end{equation}
\begin{equation}\label{curvature}
(\nabla_{a}\nabla_{b}-\nabla_{b}\nabla_{a})\omega_{d}
=-R^c{}_{dab}\,\omega_{c}-S^{c}{}_{ab}\nabla_{c}\,\omega_{d}.
\end{equation}
The tensor fields $S^{c}{}_{ab}$ and $R^c{}_{dab}$ are called the
torsion and curvature tensor fields of $\nabla_a$, respectively.
Suppose that $\{e_{\alpha}{}^a\}$ is an orthonormal frame field,
where $\alpha=0,1,2,3$. The metric components in $\{e_{\alpha}{}^a\}$ are equal to
$\eta_{\alpha\beta}=$ diag$(-1,1,1,1)$. The connection one-form of
$\nabla_a$ in $\{e_{\alpha}{}^a\}$ is defined as
\begin{equation}
\Gamma^{\alpha}{}_{\beta
a}=e^{\alpha}{}_{b}\nabla_{a}e_{\beta}{}^{b},
\end{equation}
where $\{e^{\alpha}{}_{b}\}$ is the dual frame field
of $\{e_{\alpha}{}^a\}$. The torsion and curvature two-form of $\nabla_a$
in $\{e_{\alpha}{}^a\}$ are defined as
$S^{\alpha}{}_{ab}=S^{c}{}_{ab}\,e^{\alpha}{}_{c}$ and
$R^{\alpha}{}_{\beta ab}=R^c{}_{dab}\,e^{\alpha}{}_{c}\,e_{\beta}{}^{d}$, respectively.
Then Eqs. (\ref{torsion}) and (\ref{curvature}) imply
\begin{equation}
S^{\alpha}{}_{ab}=d_a e^{\alpha}{}_b+\Gamma^{\alpha}{}_{\beta
a}\wedge e^{\beta}{}_{b},
\end{equation}
\begin{equation}
R^{\alpha}{}_{\beta ab}=d_a\Gamma^{\alpha}{}_{\beta b}+\Gamma^{\alpha}{}_{\gamma
a}\wedge \Gamma^{\gamma}{}_{\beta b},
\end{equation}
where $d_a$ is the exterior differentiation operator.

It can be shown (cf. Appendix A) that, a principal bundle ${\cal P}={\cal P}({\cal M},G)$
can be set up, where $G$ is the dS group. The Ehresmann connection $\Omega$ of ${\cal P}$
can be given (cf. Appendix B) by the geometry of the spacetime.
Moreover, an associated bundle ${\cal Q}_{M_5}={\cal Q}_{M_5}({\cal P},M_5)$ can be set up, where
$M_5$ is the 5-dimensional (5d) Minkowski space. The fiber bundle ${\cal Q}_{M_5}$
has a subbundle ${\cal Q}={\cal Q}({\cal P},D_4)$, where $D_4$ is a dS space. There is a
global section $\phi$ in ${\cal Q}$, with the help of which the gauge-invariant expressions
for the metric, torsion and curvature of the spacetime can be obtained (cf. Appendix C)
\cite{Guo76,Stelle,Lu13}:
\begin{equation}\label{metricxi}
g_{ab}=\eta_{AB}(D_{a}\xi^{A})(D_{b}\xi^{B}),
\end{equation}
\begin{equation}\label{torsionxi}
S_{cab}=\mathcal{F}_{ABab}(D_{c}\xi^{A})\xi^{B},
\end{equation}
\begin{equation}\label{R-F}
R_{cdab}-(2/l^{2})g_{a[c}g_{d]b}=\mathcal
{F}_{ABab}(D_{c}\xi^{A})(D_{d}\xi^{B}),
\end{equation}
where $A,B=0,1,2,3,4$,
$S_{cab}=g_{cd}S^d{}_{ab}$, $R_{cdab}=g_{ce}R^e{}_{dab}$,
${\cal F}_{ABab}=\eta_{AC}{\cal F}^C{}_{Bab}$,
$\xi^A=\xi^A(x)$ are the vertical coordinates of $\phi(x)$ in $\{x^\mu,\xi^A\}$,
$\{x^\mu,\xi^A\}$ is a local coordinate system on ${\cal Q}_{M_5}$, which
is given by a local section $\sigma$ in ${\cal P}$ together
with a local coordinate system $\{x^\mu\}$ ($\mu=0,1,2,3$) on ${\cal M}$
and a Minkowski coordinate system $\{\xi^A\}$ on $M_5$,
$D_a\xi^A$ is the covariant derivative (\ref{Daphi}) of $\phi$,
${\cal F}^C{}_{Bab}$ is the local curvature (\ref{Fl}) of $\Omega$, and $l$ is the radius of
$D_4$. The expressions (\ref{metricxi})--(\ref{R-F})
are gauge invariant, i.e., independent of the choice of the local
section $\sigma$ in ${\cal P}$.

Note that $\xi^A(x)$ can be interpreted as the local 5d Minkowski coordinates,
for the reason that the metric field in $D_4$ can be expressed as
$g_{ab}=\eta_{AB}(d_a\xi^A)(d_b\xi^B)$, where $\xi^A$ are coordinates in $\{\xi^A\}$.
The functions $\xi^A(x)$ are related to the local inertial coordinates $y^\mu(x)$ by
$y^\mu(x)=\sqrt{\sigma}\xi^\mu(x)$ when $\xi^4(x)\neq0$,
where $\sigma=[l/\xi^4(x)]^2$ \cite{dSSR,Guo76}.

Denote the gravitational action by $S_G$, and the action of the matter fields by
$S_M$. The total action is defined as $S=\kappa S_G+S_M$, where $\kappa$ is the coupling
constant. Suppose that $S$ is an integral over a domain ${\cal U\subseteq M}$, and
there exist tensor fields $X^{ab}$ and $Y_c{}^{ab}$, such that for any ${\cal U}$ with a
compact closure, the variation of $S$ caused by the variations of the metric and torsion
on ${\cal U}$ is equal to
\begin{equation}\label{XY}
\delta S=\int_{\cal U}\,[X^{ab}\delta g_{ab}+Y_c{}^{ab}\delta S^c{}_{ab}]\epsilon,
\end{equation}
where $\epsilon$ is a metric-compatible volume element. Then the gravitational field
equations are given by
\begin{equation}\label{XY=0}
\delta S/\delta g_{ab}\equiv X^{(ab)}=0,\quad
\delta S/\delta S^c{}_{ab}\equiv Y_c{}^{[ab]}=0.
\end{equation}
Making use of Eqs. (\ref{metricxi})--(\ref{R-F}), it can be found that
the variations of the metric and torsion are equivalent to the variations
of $\Omega^A{}_{Ba}$ and $\xi^A$, where $\Omega^A{}_{Ba}$ is the local Ehresmann
connection (\ref{omega-a}).
Note that $\delta\xi^A(x)=(\partial\xi^A/\partial y^\mu)\delta y^\mu(x)$,
where $\delta y^\mu(x)$ is arbitrary on ${\cal U}$ but $\delta\xi^A(x)$ is not.
Therefore, the gravitational field equations can also be given by
\begin{eqnarray}
\delta S/\delta\Omega^A{}_{Ba}=0,\qquad\quad\ \ \label{deltaO}\\
(\delta S/\delta\xi^A)(\partial\xi^A/\partial y^\mu)=0.\label{deltaX}
\end{eqnarray}
As we shall see, Eq. (\ref{deltaX}) is redundant. Actually, provided Eq.
(\ref{deltaO}) holds, the variation of $S$ caused by any variations of
$\Omega^A{}_{Ba}$ and $\xi^A$ on ${\cal U}$ is equal to
\begin{equation}\label{deltaS}
\delta S=\int_{\cal U}(\delta S/\delta\xi^{A})\delta\xi^{A}.
\end{equation}
Suppose that the variation is given by a group of gauge transformations, then
$\delta\xi^A=(\delta g^A{}_B)\xi^B$, where $g^{A}{}_{B}$ is a $G$-valued function,
and thus $\delta g^{A}{}_B$ is a $\mathfrak{g}$-valued function, where $\mathfrak{g}$
is the dS algebra. Since $S$ is gauge invariant, $\delta S=0$, and so \cite{Lu13},
\begin{equation}\label{deltaSx}
(\delta S/\delta\xi^{A})\xi_{B}-(\delta S/\delta\xi^{B})\xi_{A}=0.
\end{equation}
Notice that $\xi^A(x)\xi_A(x)=l^2$, then $(\partial\xi^A/\partial y^\mu)\xi_A(x)=0$,
and hence Eq. (\ref{deltaSx}) results in Eq. (\ref{deltaX}). In other words,
the gravitational field equation can be given by Eq. (\ref{deltaO}),
which is similar to the gauge field equations in the gauge theories of matter fields.
In the special gauge with $\sigma=\sigma_e$, it is easy to found that the variations
of $\Omega^A{}_{Ba}$ and $\xi^A$ are equivalent to the variations of $\Gamma^\alpha{}_{\beta a}$
and $e^\alpha{}_a$ (cf. Eqs. (\ref{phi}) and (\ref{omega-e})). As a result,
the gravitational field equations can also be given by
\begin{equation}\label{DeltaGe}
\delta S/\delta\Gamma^\alpha{}_{\beta a}=0,\quad
\delta S/\delta e^\alpha{}_a=0.
\end{equation}

The no-gravity spacetime can be defined as a spacetime where the Ehresmann connection
is flat, i.e., ${\cal F}^A{}_{Bab}=0$. The gauge-invariant expressions (\ref{torsionxi})
and (\ref{R-F}) imply that the no-gravity spacetime is locally identical with $D_4$.
It is a natural requirement that the no-gravity spacetime is a vacuum solution.
Furthermore, we would like to require that the gravitational Lagrangian density $\mathscr{L}_G$
is complete in the sense that it contains all the matrix elements
of ${\cal F}^A{}_{Bab}$, i.e., it contains both curvature and torsion (cf. Eq. (\ref{Fe})).
In the next section, we will construct a model for the dS gauge theory of gravity which
satisfies the above-mentioned requirements.

\section{Kaluza\texorpdfstring{--}{-}Klein-like model}

The Lagrangian density of the KK-like model to be constructed is defined as
$\mathscr{L}_G=\phi^*\overline{R}$, where $\overline{R}$ is the scalar curvature
of ${\cal Q}$, $\phi^*$ is the pullback map of $\phi$.
The scalar curvature $\overline{R}$
is determined by the metric and torsion of ${\cal Q}$.
The metric of ${\cal Q}$ is defined as
\begin{equation}\label{gbar}
\overline{g}=g_{\mu\nu}dx^{\mu}\otimes
dx^{\nu}+\alpha\eta_{AB}\theta^{A}\otimes\theta^{B},
\end{equation}
where $x^\mu$ are coordinates in $\{x^\mu,\xi^A\}$, $g_{\mu\nu}$ are components of
$g_{ab}$ in $\{x^\mu\}$, $\otimes$ stands for a tensor product,
$\theta^A$ are several vertical one-form fields (\ref{theta}), and
$\alpha$ is a dimensionless constant. Since $\theta^A$ are gauge covariant,
Eq. (\ref{gbar}) is gauge invariant. The first part of Eq. (\ref{gbar}) comes from $g_{ab}$,
and the second part comes from the metric of $D_4$. The constant $\alpha=1$
in Ref. \cite{Lu13}, while it is to be determined here.
With the help of Eqs. (\ref{metricxi}) and (\ref{theta}), it can be shown that \cite{Lu13}
\begin{equation}
[\phi^{*}(g_{\mu\nu}dx^{\mu}\otimes dx^{\nu})]_{ab}=
[\phi^{*}(\eta_{AB}\theta^{A}\otimes\theta^{B})]_{ab}=g_{ab}.
\end{equation}
Moreover, the torsion of ${\cal Q}$ can be defined as \cite{Lu13}:
\begin{equation}
\overline{S}=S^\sigma{}_{\mu\nu}X_\sigma\otimes dx^\mu\otimes dx^\nu,
\end{equation}
where $S^\sigma{}_{\mu\nu}$ are components of $S^c{}_{ab}$ in
$\{x^\mu\}$, and $X_\sigma$ are horizontal lifts (\ref{Xmu}) of the coordinate
basis vector fields of $\{x^\mu\}$. Because $D_4$ is torsion free, $\overline{S}$ only contains
the spacetime torsion. Note that $X_\sigma$ are gauge invariant, and so $\overline{S}$
is gauge invariant. A local frame field $\{\overline{E}_{\cal A}\}=\{X_\mu,E_\alpha\}$ can be
defined on ${\cal Q}$, with $\{\overline{E}^{\cal A}\}=\{dx^\mu,E^\alpha\}$ as its dual,
where ${\cal A}=0,1,\cdots,7$, and, $E_\alpha$ and $E^\alpha$ are several vertical
vector fields and one-form fields, respectively (cf. Appendix C).
The connection coefficients and curvature components of ${\cal Q}$ in
$\{\overline{E}_{\cal A}\}$ can be calculated by the following formulae \cite{Cho}:
\begin{equation}
\overline{\Gamma}^{\cal C}{}_{\cal AB}=\frac{1}{2}\,\overline{g}^{\,\cal CD}
[\overline{E}_{\cal A}(\overline{g}_{\cal BD})+\overline{E}_{\cal B}
(\overline{g}_{\cal AD})-\overline{E}_{\cal D}(\overline{g}_{\cal AB})]
-\overline{K}^{\cal C}{}_{\cal AB}-\frac{1}{2}(\overline{C}^{\cal C}{}_{\cal AB}
+\overline{C}_{\cal AB}{}^{\cal C}+\overline{C}_{\cal BA}{}^{\cal C}),
\end{equation}
\begin{equation}
\overline{R}^{\cal C}{}_{\cal DAB}=2\overline{E}_{[\cal A}
(\overline{\Gamma}^{\cal C}{}_{|\cal D|B]})+2\overline{\Gamma}^{\cal C}{}_{\cal E[A}
\overline{\Gamma}^{\cal E}{}_{|\cal D|B]}-\overline{\Gamma}^{\cal C}{}_{\cal DE}
\overline{C}^{\cal E}{}_{\cal AB},
\end{equation}
where $\overline{K}^{\cal C}{}_{\cal AB}$ are contorsion components
related to the torsion components $\overline{S}^{\,\cal C}{}_{\cal AB}$ by
\begin{equation}
\overline{K}^{\cal C}{}_{\cal AB}=\frac{1}{2}(\overline{S}^{\,\cal C}{}_{\cal AB}
+\overline{S}_{\cal AB}{}^{\cal C}+\overline{S}_{\cal BA}{}^{\cal C}),
\end{equation}
and $\overline{C}^{\cal C}{}_{\cal AB}$ are defined by the commutation relations:
\begin{equation}\label{commutation}
[\overline{E}_{\cal A},\overline{E}_{\cal B}]
=\overline{C}^{\cal C}{}_{\cal AB}\overline{E}_{\cal C}.
\end{equation}
It can be shown that \cite{Lu13}
\begin{eqnarray}
\overline{C}^{\sigma}{}_{\mu\nu}=0,~\overline{C}^{\alpha}{}_{\mu\nu}=-\mathcal
{F}^{A}{}_{B\mu\nu}\xi^{B}E_{A}{}^{\alpha},\qquad \qquad \quad\,\label{C1}\\
\overline{C}^{\nu}{}_{\mu\alpha}=0,~\overline{C}^{\beta}{}_{\mu\alpha}=[X_{\mu}(E_{\alpha}{}^{A})
+\Omega^{A}{}_{B\mu}E_{\alpha}{}^{B}]E_{A}{}^{\beta},\label{C2}\\
\overline{C}^{\mu}{}_{\alpha\beta}=0,~\overline{C}^{\,\gamma}{}_{\alpha\beta}=
C^{\gamma}{}_{\alpha\beta},\qquad \qquad \qquad \qquad\quad\ \,
\end{eqnarray}
where $\overline{C}^{\alpha}{}_{\mu\nu}$, etc. are short for
$\overline{C}^{4+\alpha}{}_{\mu\nu}$, etc., $\Omega^A{}_{B\mu}$ and ${\cal F}^{A}{}_{B\mu\nu}$
are components of $\Omega^A{}_{Ba}$ and ${\cal F}^A{}_{Bab}$ in $\{x^\mu\}$,
$\xi^B$ are coordinates in $\{x^\mu,\xi^A\}$, $E_A{}^\alpha$ and $E_\alpha{}^A$ are some functions
related to $E_\alpha$ (cf. Eqs. (\ref{Ealpha}) and (\ref{tpartialA})),
and $C^{\gamma}{}_{\alpha\beta}$ are determined by the commutation relations of $\{E_\alpha\}$.
Moreover, it can be checked that
\begin{eqnarray}
\overline{\Gamma}^{\sigma}{}_{\mu\nu}=\Gamma^{\sigma}{}_{\mu\nu},\,
\overline{\Gamma}^{\gamma}{}_{\alpha\beta}=(\Gamma_F)^{\gamma}{}_{\alpha\beta},\qquad\ \ \\
\overline{\Gamma}^{\alpha}{}_{\mu\nu}=\frac{1}{2}\mathcal
{F}^{A}{}_{B\mu\nu}\xi^{B}E_{A}{}^{\alpha},\qquad \qquad \quad\ \,\label{gammabar}\\
\overline{\Gamma}^{\mu}{}_{\nu\alpha}=\overline{\Gamma}^{\mu}{}_{\alpha\nu}
=\frac{1}{2}\alpha\mathcal{F}_{AB\nu}{}^{\mu}\xi^{B}E_{\alpha}{}^{A},\quad~\,\\
\overline{\Gamma}^{\mu}{}_{\alpha\beta}=0,\,
\overline{\Gamma}^{\beta}{}_{\mu\alpha}=0,\,
\overline{\Gamma}^{\beta}{}_{\alpha\mu}=-\overline{C}^{\beta}{}_{\alpha\mu},\label{gammabar2}
\end{eqnarray}
where $\Gamma^\sigma{}_{\mu\nu}$ are connection coefficients
of ${\cal M}$ in $\{x^\mu\}$, $(\Gamma_F)^{\gamma}{}_{\alpha\beta}$ are induced by
the connection coefficients of $D_4$ in a frame field $\{E_\alpha{}^a\}$ related to
$\{E_\alpha\}$ (cf. Appendix C), and ${\cal F}^A{}_{B\nu}{}^\mu=g^{\mu\sigma}{\cal F}^A{}_{B\nu\sigma}$.
Furthermore, it can be checked that
\begin{eqnarray}
\overline{R}^{\mu\nu}{}_{\mu\nu}=R-\frac{3}{4}\alpha\mathcal
{F}_{AB}{}^{\mu\nu}\mathcal {F}^{A}{}_{C\mu\nu}\xi^{B}\xi^{C},\ \ \,\\
\overline{R}^{\alpha\beta}{}_{\alpha\beta}=\alpha^{-1}R_F,
\qquad \qquad \qquad \qquad \ \ \,\label{Rbar1}\\
\overline{R}^{\mu\alpha}{}_{\mu\alpha}=\frac{1}{4}\alpha\mathcal
{F}_{AB}{}^{\mu\nu}\mathcal {F}^{A}{}_{C\mu\nu}\xi^{B}\xi^{C},
\qquad \quad\label{Rbar2}
\end{eqnarray}
where $\overline{R}^{\mu\nu}{}_{\mu\nu}=g^{\nu\sigma}\overline{R}^{\mu}{}_{\sigma\mu\nu}$,
$\overline{R}^{\alpha\beta}{}_{\alpha\beta}=
\overline{g}^{\beta\gamma}\overline{R}^{\alpha}{}_{\gamma\alpha\beta}$,
$\overline{R}^{\mu\alpha}{}_{\mu\alpha}=
\overline{g}^{\alpha\beta}\overline{R}^{\mu}{}_{\beta\mu\alpha}$,
$R=g^{db}R^a{}_{dab}$ is the scalar curvature of ${\cal M}$,
and $R_F=4\Lambda$ is the scalar curvature of $D_4$, where
$\Lambda=3/l^2$. Therefore, the scalar curvature of ${\cal Q}$ is equal to
\begin{eqnarray}\label{Rbar}
\overline{R}=R+\alpha^{-1}R_F-(1/4)\alpha\mathcal {F}_{AB}{}^{\mu\nu}\mathcal
{F}^{A}{}_{C\mu\nu}\xi^{B}\xi^{C},
\end{eqnarray}
and hence
\begin{eqnarray}
\phi^*\overline{R}=R+4\Lambda/\alpha-(1/4)\alpha|S|^2,
\end{eqnarray}
where $|S|^2=S^{abc}S_{abc}$, and Eq. (\ref{Fe}) has been used.
In order to make the no-gravity spacetime be a vacuum solution, $4\Lambda/\alpha$
should be equal to $-2\Lambda$. As a result, $\alpha=-2$, and the Lagrangian density
of the KK-like model being constructed is
\begin{equation}\label{Lagrangian}
\mathscr{L}_G=\phi^{*}\overline{R}=R-2\Lambda+(1/2)|S|^2.
\end{equation}

For generality, we will present the gravitational field equations for
a group of models:
\begin{equation}\label{Lagrangian2}
\mathscr{L}_G=R-2\Lambda+\beta|S|^2,
\end{equation}
where $\beta$ is a dimensionless parameter. For the EC theory, $\beta=0$;
for the KK-like model in Ref. \cite{Lu13}, $\beta=-1/4$;
for the KK-like model (\ref{Lagrangian}), $\beta=1/2$. The gravitational
field equations of the models (\ref{Lagrangian2}) are as follows:
\begin{eqnarray}\label{1stEq}
R_{ba}-\frac1 2Rg_{ab}+\Lambda g_{ab}-2\beta\nabla_{c}S_{ab}{}^{c}
-\beta S_{acd}T_{b}{}^{cd}-\frac1 2\beta|S|^2g_{ab}
+2\beta S^{cd}{}_aS_{cdb}=\frac{1}{2\kappa}\Sigma_{ab},
\end{eqnarray}
\begin{equation}\label{2ndEq}
 T^{a}{}_{bc}-4\beta S_{[bc]}{}^{a}=-\frac{1}{\kappa}\tau_{bc}{}^{a},
\end{equation}
where $R_{ba}=R^c{}_{bca}$, $T^a{}_{bc}=S^a{}_{bc}+2\delta^a{}_{[b}S_{c]}$,
$\delta^a{}_b$ is the Kronecker delta, $S_c=S^b{}_{cb}$, $\Sigma_{ab}=\Sigma_\alpha{}^ce^\alpha{}_ag_{bc}$
is the canonical energy-momentum tensor, where $\Sigma_\alpha{}^c=\delta S_M/\delta e^\alpha{}_c$,
and, $\tau_{bc}{}^a=\tau_\alpha{}^{\beta a}e^\alpha{}_be_{\beta c}$ is the spin tensor,
where $\tau_\alpha{}^{\beta a}=\delta S_M/\delta\Gamma^\alpha{}_{\beta a}$,
and $e_{\beta c}=\eta_{\alpha\beta}e^\alpha{}_c$.
Note that the cases with $\beta=1,1/4$ or $-1/2$ are not so general
because they can only describe the matter fields with $\tau_b\equiv\tau_{bc}{}^c=0$,
$\tau_{[abc]}=0$, or $g_{a[b}\tau_{c]}-\tau_{a[bc]}+\tau_{bca}=0$, respectively.
For the cases with $\beta\neq1,1/4,-1/2$, the solution
of Eq. (\ref{2ndEq}) is
\begin{equation}\label{tor-spin}
S_{abc}=(8\beta^2+2\beta-1)^{-1}[-2(4\beta-1)g_{a[b}S_{c]}
-(2\beta-1)(1/\kappa)\tau_{bca}+(4\beta/\kappa)\tau_{a[bc]}],
\end{equation}
where $S_c=[1/2\kappa(1-\beta)]\tau_c$. The symmetric energy-momentum tensor
$T_{ab}=-2\delta S_M/\delta g^{ab}$ is related to
$\Sigma_{ab}$ and $\tau_{bc}{}^a$ by
\begin{eqnarray}\label{T-Sigma}
T_{ab}&=&\Sigma_{ab}+(\nabla_c+S_c)(\tau_{ab}{}^c-\tau_a{}^c{}_b+\tau^c{}_{ba})\nonumber\\
&=&\Sigma_{(ab)}+2(\nabla_c+S_c)\tau^c{}_{(ab)}.
\end{eqnarray}
Hence, if the spin tensor is equal to zero, then $S^c{}_{ab}=0$,
$\Sigma_{ab}=T_{ab}$, and Eq. (\ref{1stEq}) reduces to the Einstein
equation with a cosmological constant
when $1/2\kappa=8\pi$, i.e., $\kappa=1/16\pi$. Generally, Eq. (\ref{1stEq})
can be expressed as an Einstein-like equation:
\begin{equation}\label{Elike}
\mathring R_{ab}-\frac1 2\mathring Rg_{ab}+\Lambda g_{ab}
=\frac1{2\kappa}(T_{\rm eff})_{ab},
\end{equation}
where $\mathring R=g^{ab}\mathring R_{ab}$, $\mathring R_{ab}=\mathring R^c{}_{acb}$,
$\mathring R^c{}_{dab}$ is the torsion-free curvature,
and $(T_{\rm eff})_{ab}$ is the effective energy-momentum tensor which satisfies
\begin{eqnarray}\label{Teff}
\frac1{2\kappa}(T_{\rm eff})_{ab}=\frac1{2\kappa}T_{ab}
-\frac1\kappa(\frac12S_{(a}{}^{cd}\tau_{|cd|b)}+K^d{}_{(b|c|}\tau^c{}_{a)d})
+\frac12(1+2\beta)S^{cd}{}_{(a}S_{b)cd}\nonumber\\
-S_{(ab)}{}^cS_c-\frac14(T_{cde}K^{dec}-2\beta|S|^2)g_{ab}
-\frac14S_{acd}S_b{}^{cd}-2\beta S_{cda}S^{[cd]}{}_b,
\end{eqnarray}
where $K^d{}_{bc}$ is the spacetime contorsion.

\section{Cosmological meaning}
Let us assume that the matter fields in the universe
can be described by a spin fluid \cite{Kuchowicz}
with the energy-momentum tensor and spin tensor being
\begin{equation}\label{T}
T_{ab}=\rho U_aU_b+p(g_{ab}+U_aU_b),
\end{equation}
\begin{equation}\label{tau}
\tau_{bc}{}^a=\tau_{bc}U^a,
\end{equation}
where $\rho$ is the rest energy density, $p$ is the hydrostatic pressure,
$U^a$ is the four-velocity of the fluid particles, and $\tau_{bc}$ is the spin density
two-form which satisfies $\tau_{bc}U^c=0$. This spin fluid is one of the
extensions \cite{Hehl74,Kuchowicz} of the special relativistic Weyssenhoff fluid \cite{Weyssenhoff}.
Substitution of Eq. (\ref{tau}) into Eq. (\ref{Teff}) yields
\begin{eqnarray}\label{Teff2}
\frac1{2\kappa}(T_{\rm eff})_{ab}=\frac1{2\kappa}T_{ab}
+\frac1{\kappa^2}(8\beta^2+2\beta-1)^{-2}
[(\beta-\frac12)(8\beta^2+2\beta-1)\tau_{ac}\tau_b{}^c\nonumber\\
+(-4\beta^3-3\beta^2+\frac14)2s^2U_aU_b
+(-6\beta^3-\frac12\beta^2+\beta-\frac18)2s^2g_{ab}],
\end{eqnarray}
where $s^2=\tau_{bc}\tau^{bc}/2$ is the spin density squire.
Furthermore, the metric field of the universe
is supposed to be an RW metric with the line element
\begin{equation}\label{RW}
ds^2=-dt^2+a^2(t)[\frac{dr^2}{1-kr^2}+r^2(d\theta^2+\sin^2\theta d\varphi^2)],
\end{equation}
where $t$ is the cosmic time with
$(\partial/\partial t)^a=U^a$, and $k=0,\pm1$. According to Eq. (\ref{RW}),
the left-hand side of the Einstein-like equation (\ref{Elike}) is diagonal in the coordinate
system $\{t,r,\theta,\varphi\}$, then $(T_{\rm eff})_{ab}$
should be diagonal too, and so the term proportional to $\tau_{ac}\tau_b{}^c$ in Eq.
(\ref{Teff2}) should be diagonal, which implies that $\beta=1/2$ or $\tau_{bc}=0$.
In other words, among the group of models (\ref{Lagrangian2}), only the
KK-like model (\ref{Lagrangian}) has nonzero torsion effect in the
present case.

It should be remarked that the torsion field given by Eqs. (\ref{tor-spin})
and (\ref{tau}) is not homogenous and isotropic in the usual sense, i.e.,
not invariant under the isometric transformations of the cosmic space.
However, we may still think that the torsion field is homogeneous and isotropic,
for the reason that it is algebraically related to the spin tensor, which can be assumed to be
homogeneous in the sense that the spin density $s$ is a function of the
cosmic time, and isotropic in the sense that the spin orientations given
by $\tau_{bc}$ are random at the cosmic scale.

In fact, for the KK-like model (\ref{Lagrangian}), Eq. (\ref{Teff2}) becomes
\begin{equation}
(T_{\rm eff})_{ab}=T_{ab}-\frac1{2\kappa}s^2U_aU_b
-\frac1{2\kappa}s^2(g_{ab}+U_aU_b),
\end{equation}
which indicates that the torsion effect is equivalent to an ideal
fluid with the rest energy density being negative, and equal to the hydrostatic pressure.
It is easy to show that this torsion effect may replace the big-bang singularity
by a big bounce \cite{Kuchowicz2,Poplawski}.

\section{Remarks}

In this paper, a new KK-like model (\ref{Lagrangian}) of the dS gauge theory of
gravity is constructed from the geometry related to the gauge-invariant expressions
(\ref{metricxi})--(\ref{R-F}) of the metric, torsion and curvature.
The model is complete in the sense that the Lagrangian density contains both
curvature and torsion. The model is simple in the sense that the Lagrangian density
only contains a curvature term, a torsion term and a constant term.

It is remarkable that the coefficient of the torsion term in the Lagrangian density
(\ref{Lagrangian}) is determined by the requirement that the no-gravity spacetime is a
vacuum solution. This coefficient is significant since it gives the only one model of
such type that may avoid the big-bang singularity of the RW universe
filled with a spin fluid (\ref{T})(\ref{tau}).

\section*{Acknowledgments}
I would like to thank the late Prof. H.-Y. Guo, and Profs. C.-G.
Huang, Z.-B. Li, X.-P. Zhu, X.-W. Liu, S.-D. Liang and T. Harko for their help
and some useful discussions.

\appendix

\section{Firber bundles}

Suppose that ${\cal P}_H={\cal P}_H({\cal M}, H)$ is an orthonormal frame bundle,
where $H$ is the Lorentz group. We can identify $H$ to a
subgroup of $G$ by identifying $h\in H$ to
\begin{equation}\label{hAB}
\left(
\begin{array}{cc}
h&0\\0&1
\end{array}
\right)\in G.
\end{equation}
Then $H$ can act on $G$ by the group multiplication, and we can set up
an associated bundle ${\cal Q}_G={\cal Q}_G({\cal P}_H,G)$, which turns
out to be a principal bundle ${\cal P}={\cal P}({\cal M},G)$ \cite{Wise},
with the group action given by
\begin{equation}
(p_{H}\cdot g)g_1=p_{H}\cdot(gg_1),
\end{equation}
where $p_H\in{\cal P}_H$, $g,g_1\in G$ and
\begin{equation}
p_{H}\cdot g=\{(p_Hh^{-1},hg)|h\in H\}\in{\cal Q}_G={\cal P}.
\end{equation}
Likewise, we can set up an associated bundle ${\cal Q}_{GL}={\cal Q}_{GL}
({\cal P}_H,GL(5,\mathbb{R}))$, which turns out to be a principal bundle
${\cal P}_{GL}={\cal P}_{GL}({\cal M}, GL(5,\mathbb{R}))$,where $GL(5,\mathbb{R})$
is a general linear group. Since $G$ is a subgroup of $GL(5,\mathbb{R})$,
${\cal P}$ is a subbundle of ${\cal P}_{GL}$.

Note that $G$ can act on $M_5$ and $D_4$, once a Minkowski coordinate
system $\{\xi^A\}$ is chosen. Then we can set up two associated bundles ${\cal Q}_{M_5}
={\cal Q}_{M_5}({\cal P},M_5)$ and ${\cal Q}={\cal Q}({\cal P},D_4)$,
such that ${\cal Q}$ is a subbundle of ${\cal Q}_{M_5}$.
Suppose that $\sigma$ is a local section in ${\cal P}$. Then $\sigma$
together with $\{x^\mu\}$ and $\{\xi^A\}$ presents a coordinate system $\{x^{\mu},\xi^{A}\}$
on ${\cal Q}_{M_5}$. Let $\sigma'$ be another local section in ${\cal P}$,
and $g(x)$ be the transition function, such that
\begin{equation}\label{gaugeT}
\sigma'(x)=\sigma(x)[g(x)]^{-1}.
\end{equation}
The transformation (\ref{gaugeT}) of $\sigma$ is called the gauge transformation,
which induces a coordinate transformation of ${\cal Q}_{M_5}$:
\begin{equation}\label{xitran}
\xi'^A=[g(x)]^A{}_B\xi^B.
\end{equation}

For any point $x_0\in{\cal M}$, there exists an orthonormal frame field $\{e_{\alpha}{}^a\}$
defined on a domain ${\cal O}\subseteq{\cal M}$, such that $x_0\in{\cal O}$.
The frame field $\{e_{\alpha}{}^a\}$ presents a local section
$\sigma_{H}(x)=(x,\{e_{\alpha}{}^a|_x\})$ in ${\cal P}_{H}$,
which induces a local section $\sigma_e(x)=\sigma_{H}(x)\cdot e$ in
${\cal P}$, where $e$ stands for the identity element of $G$.
Assume that $\mathring{\xi}\in D_4$, whose coordinates
in $\{\xi^A\}$ are $\mathring{\xi}^A=(0,0,0,0,l)^{T}$.
Then $\sigma_e(x)\cdot\mathring{\xi}$ is a local section in
${\cal Q}$, whose value at $x_0$ is independent of the choice of
the frame field $\{e_{\alpha}{}^a\}$, as will be shown below.

Suppose that $\{e'_{\alpha}{}^a\}$ is another orthonormal frame field defined on
a domain ${\cal O'}\subseteq{\cal M}$, such that $x_0\in{\cal O'}$.
Then there exists a function $h(x)$, which is defined on
${\cal O}\cap{\cal O'}$, and valued on $H$, such that
$e'_{\alpha}{}^a|_x=e_{\beta}{}^a|_x\,h^{\beta}{}_{\alpha}(x)$ for any $x\in{\cal O}\cap{\cal O'}$.
The frame field $\{e'_{\alpha}{}^a\}$ presents a local section
$\sigma'_{H}(x)=(x,\{e'_{\alpha}{}^a|_x\})$ in ${\cal P}_{H}$,
which is equal to $\sigma_H(x)h(x)$ on ${\cal O}\cap{\cal O'}$.
The local section $\sigma'_{H}(x)$ induces a local section
$\sigma'_e(x)=\sigma'_{H}(x)\cdot e$ in ${\cal P}$,
which is equal to $\sigma_e(x)h(x)$ on ${\cal O}\cap{\cal O'}$.
The local section $\sigma'_e(x)$ together with $\mathring{\xi}$
gives a local section $\sigma'_e(x)\cdot\mathring{\xi}$ in ${\cal Q}$, which is equal to
\[
\sigma_e(x)h(x)\cdot\mathring{\xi}=
\sigma_e(x)\cdot h(x)\mathring{\xi}=\sigma_e(x)\cdot\mathring{\xi},
\]
on ${\cal O}\cap{\cal O'}$.
As a result, $\sigma_e(x_0)\cdot\mathring{\xi}$ is independent of the choice of
the frame field $\{e_{\alpha}{}^a\}$, and a global section $\phi$ of ${\cal Q}$
can be defined by \cite{Lu13}
\begin{equation}\label{phi}
\phi(x_0)=\sigma_e(x_0)\cdot\mathring{\xi}.
\end{equation}

\section{Ehresmann connection}

Suppose that at any $p\in{\cal P}$, the tangent space of ${\cal P}$ can be decomposed into a direct sum:
\begin{equation}\label{Dsum}
T_p{\cal P}=V_p\oplus H_p,
\end{equation}
where $V_p$ is the vertical subspace defined as the tangent space of fiber,
and, $H_p$ is called the horizontal subspace, and satisfies
\begin{equation}\label{Hpg}
H_{pg}=R_{g*}[H_p]
\end{equation}
for any $g\in G$, where $R_g$ is the group action of $G$ on ${\cal P}$, and $R_{g*}$ denotes
the pushforward map of $R_g$.
The decomposition (\ref{Dsum}) can be used to define a $\mathfrak{g}$-valued one-form
$\Omega$ on ${\cal P}$, called the Ehresmann connection, such that for any $X\in T_p{\cal P}$,
\begin{equation}\label{Omega}
\Omega(X)=(R_p^{-1})_*X^V,
\end{equation}
where $X^V\in V_p$ is the vertical part of $X$ defined by Eq. (\ref{Dsum}),
and $R_p$ is a diffeomorphism defined by $R_p\,g=pg$.
The condition (\ref{Hpg}) implies that
\begin{equation}\label{RgOmega}
R_g^{\,*}\,\Omega=I_{g^{-1}*}\circ\Omega,
\end{equation}
where $\circ$ denotes a composition, and $I_g$ is a Lie group isomorphism
defined by $I_g\,g_1=g\,g_1g^{-1}$ for any $g_1\in G$. Conversely, given the Ehresmann
connection $\Omega$ as a $\mathfrak{g}$-valued one-form which satisfies
Eq. (\ref{RgOmega}), we can give a direct sum decomposition (\ref{Dsum}) which
is compatible with Eq. (\ref{Omega}), in the following way.
For any $p\in{\cal P}$ and $X\in T_p\,{\cal P}$, define $X^V=R_{p\,*}[\Omega(X)]$, $X^H=X-X^V$,
and $H_p={\rm Ker}(\Omega|_p)$. Then $X=X^V+X^H$, $X^V\in V_p$, $X^H\in H_p$,
and Eq. (\ref{Hpg}) holds.

The local Ehresmann connection with respect to $\sigma$ is defined as
\begin{equation}\label{omega-a}
\Omega_a=(\sigma^*\Omega)_a,
\end{equation}
which is a $\mathfrak{g}$-valued one-form on ${\cal U}$, where ${\cal U}$
is the definition domain of $\sigma$.
Under the gauge transformation (\ref{gaugeT}), $\Omega_a$ transforms in the
following way:
\begin{equation}\label{gaugeO}
\Omega'^A{}_{Ba}=[g(x)]^A{}_C\,\Omega^C{}_{Da}[g(x)^{-1}]^D{}_B
+[g(x)]^A{}_C\,d_a[g(x)^{-1}]^C{}_B,
\end{equation}
where $A,B$, etc. are matrix indices, and $\Omega'_a=(\sigma'^*\Omega)_a$.
Conversely, given $\sigma$ and $\Omega_a$ as a
$\mathfrak{g}$-valued one-form on ${\cal U}$, we can construct an Ehresmann
connection $\Omega$ on $\pi^{-1}[{\cal U}]$, where $\pi$ is the projection map
of ${\cal P}$, such that Eq. (\ref{omega-a}) holds, in the following way.
For any $p\in\pi^{-1}[{\cal U}]$ and $X\in T_p\,{\cal P}$, define
\begin{equation}\label{omegafl}
\Omega|_p(X)=I_{g^{-1}*}
[R_{\sigma(x)*}^{-1}(R_{g^{-1}*}X-\sigma_*\pi_*X)
+\Omega_a|_x(\pi_*X)^a],
\end{equation}
where $x=\pi(p)$, and $g$ is determined by $p=\sigma(x)g$. Likewise,
given another local section $\sigma'$ defined on a domain ${\cal U}'$, and $\Omega'_a$ as
a $\mathfrak{g}$-valued one-form on ${\cal U}'$, we can construct an Ehresmann connection
$\Omega'$ on $\pi^{-1}[{\cal U}']$, such that $\Omega'_a=(\sigma'^*\Omega)_a$.
 If $\Omega_a$ and $\Omega'_a$
satisfy the relation (\ref{gaugeO}), then $\Omega$ and $\Omega'$
given by them are the same on $\pi^{-1}[{\cal U}\cap{\cal U'}]$.

The curvature of $\Omega$ is defined as ${\cal F}=D{\Omega}$, where $D\Omega$
is a $\mathfrak{g}$-valued two-form on ${\cal P}$, such that for any vector fields
$X_1$ and $X_2$ on ${\cal P}$, $D\Omega(X_1,X_2)=(d\Omega)(X_1^H,X_2^H)$.
The local curvature of $\Omega$ with respect to $\sigma$ is defined by
\begin{equation}\label{Fl}
{\cal F}_{ab}=(\sigma^*{\cal F})_{ab},
\end{equation}
which is a $\mathfrak{g}$-valued two-form on ${\cal U}$.
It can be shown that
\begin{equation}\label{Flmatrix}
{\cal F}^A{}_{Bab}=d_a\Omega^A{}_{Bb}+\Omega^A{}_{Ca}\wedge\Omega^C{}_{Bb}.
\end{equation}
Under the gauge transformation (\ref{gaugeT}), ${\cal F}_{ab}$ transforms as
\begin{equation}\label{FT}
{\cal F}'^A{}_{Bab}=[g(x)]^A{}_C\,{\cal F}^{\,C}{}_{Dab}\,[g(x)^{-1}]^D{}_B.
\end{equation}

In fact, $\Omega$ can be given by the RC geometry of ${\cal M}$. An orthonormal frame field
$\{e_{\alpha}{}^a\}$ defined on a domain ${\cal O}\subseteq{\cal M}$ presents a local section
$\sigma_e=(x,\{e_{\alpha}{}^a|_x\})\cdot e$ in ${\cal P}$. The local Ehresmann connection with
respect to $\sigma_e$ can be defined as \cite{Lu13}
\begin{equation}\label{omega-e}
(\Omega_e)_a=\left(
\begin{array}{cc}
\Gamma^{\alpha}{}_{\beta a}&l^{-1}e^{\alpha}{}_{a}\\
-l^{-1}e_{\beta a}&0
\end{array}
\right).
\end{equation}
Suppose that $\{e'_{\alpha}{}^a\}$ is another orthonormal
frame field defined on a domain ${\cal O'}\subseteq{\cal M}$,
and ${\cal O}\cap{\cal O'}$ is nonempty.
Then there exists a function $h(x)$ which is defined on ${\cal O}\cap{\cal O'}$
and valued on $H$, such that $e'_\alpha{}^a|_x=e_\beta{}^a|_x\,h^\beta{}_\alpha(x)$
for any $x\in{\cal O}\cap{\cal O'}$. The frame field $\{e'_{\alpha}{}^a\}$
presents another local section $\sigma'_e=(x,\{e'_{\alpha}{}^a|_x\})\cdot e$ in ${\cal P}$.
The local Ehresmann connection $(\Omega'_e)_a$ with respect to $\sigma'_e$
can be defined in the same way as Eq. (\ref{omega-e}).
Then $(\Omega_e)_a$ and $(\Omega'_e)_a$ satisfy the relation (\ref{gaugeO}) with
\begin{equation}
g(x)=
\left(
\begin{array}{cc}
h(x)^{-1}&0\\0&1
\end{array}
\right).
\end{equation}
Hence, $(\Omega_e)_a$ and $(\Omega'_e)_a$ give the same
Ehresmann connection on $\pi^{-1}[{\cal O}\cap{\cal O'}]$.
By this way, an Ehresmann connection $\Omega$ of ${\cal P}$ can be well defined.
For some interesting interpretations of Eq. (\ref{omega-e}),
one may refer to Refs. \cite{Guo76,Guo07,Tseytlin,Wise}.
According to Eq. (\ref{Flmatrix}), the local curvature of $\Omega$ with respect to
$\sigma_e$ is equal to
\begin{equation}\label{Fe}
({\cal F}_e)_{ab}=\left(
\begin{array}{cc}
R^{\alpha}{}_{\beta ab}-l^{-2}e^{\alpha}{}_{a}\wedge e_{\beta b}
&l^{-1}S^{\alpha}{}_{ab}\\
-l^{-1}S_{\beta ab}&0
\end{array}
\right).
\end{equation}

\section{Horizontal and vertical fields}

Let $v^a$ be a vector at ${x_0\in\cal M}$. Its horizontal lift in ${\cal P}$
is a vector field $v_1$ on $\pi^{-1}(x_0)$, which satisfies
$v_1|_p\in H_p$ for any $p\in\pi^{-1}(x_0)$, and $(\pi_*v_1)^a=v^a$.
Note that the matrix elements present a natural coordinate system $\{m^A{}_B\}$
on $GL(5,\mathbb{R})$, and a local section $\sigma$ in ${\cal P}$ together with $\{x^\mu\}$
and $\{m^A{}_B\}$ presents a local coordinate system $\{x^\mu,m^A{}_B\}$ on ${\cal P}_{GL}$.
It can be shown that
\begin{equation}
v_1=v^\mu\partial_\mu-\Omega^A{}_{Ca}v^ag^C{}_B\,\partial/\partial\,m^A{}_B,
\end{equation}
where $v^\mu$ are components of $v^a$ in $\{x^\mu\}$, $\partial_\mu=\partial/\partial x^\mu$
and $\partial/\partial\,m^A{}_B$ are the coordinate basis vector fields of $\{x^\mu,m^A{}_B\}$,
and $g^C{}_B=m^C{}_B$. The horizontal lift $\overline{v}$ of $v^a$ in ${\cal Q}$
is a vector field on $\{p\cdot\xi_1\,|\,p\in\pi^{-1}(x_0),\xi_1\in D_4\}$, whose value at
$p\cdot\xi_1$ is the pushforward of $v_1|_p$ by the map $p_1\rightarrow p_1\cdot\xi_1$,
where $p_1\in{\cal P}$. It can be shown that
\begin{equation}\label{vbar}
\overline{v}=v^\mu\partial_\mu-\Omega^A{}_{Ba}v^a\xi^B\partial_A,
\end{equation}
where $\partial_\mu$ and $\partial_A$ are the coordinate basis vector fields of $\{x^\mu,\xi^A\}$.
The covariant derivative $D_a\phi$ of the global section $\phi$ is defined as follows.
At any $x\in{\cal M}$, $D_a\phi$ is a $V_{\phi(x)}$-valued one-form, where $V_{\phi(x)}$ is the tangent
space of fiber, such that for any $v^a\in T_x\,{\cal M}$,
\begin{equation}
v^a(D_a\phi)=\phi_*v^a-\overline{v}|_{\phi(x)}.
\end{equation}
It can be verified that \cite{Lu13}
\begin{equation}\label{Daphi}
D_a\phi=(D_a\xi^A)\partial_A|_{\phi(x)}=(d_a\xi^A+\Omega^A{}_{Ba}\xi^B)\partial_A|_{\phi(x)},
\end{equation}
where $\xi^A$ are short for $\xi^A(x)\equiv\xi^A[\phi(x)]$. In the special gauge with $\sigma=\sigma_e$,
$\xi^A(x)$ and $D_a\xi^A$ are identical to $(0,0,0,0,l)^T$ and
$(e^0{}_a,e^1{}_a,e^2{}_a,e^3{}_a,0)^T$.
Under the gauge transformation (\ref{gaugeT}), $\xi^A(x)$ transforms as
\begin{equation}\label{xitran2}
\xi'^A(x)=[g(x)]^A{}_B\,\xi^B(x),
\end{equation}
and $D_a\xi^A$ transforms as
\begin{equation}\label{Datran}
D_a\xi'^A=[g(x)]^A{}_BD_a\xi^B.
\end{equation}
Making use of Eqs. (\ref{FT}), (\ref{Fe}), (\ref{xitran2}) and (\ref{Datran}),
it is easy to verify the gauge-invariant expressions (\ref{metricxi})--(\ref{R-F})
for the metric, torsion and curvature.

Denote the coordinate basis vector fields of $\{x^\mu\}$ by $(\partial_\mu)^a$.
It follows from Eq. (\ref{vbar}) that the horizontal lifts of $(\partial_\mu)^a$
on ${\cal Q}$ are \cite{Lu13}
\begin{equation}\label{Xmu}
X_{\mu}=\partial_{\mu}-\Omega^{A}{}_{B\mu}\xi^{B}\partial_{A}.
\end{equation}
Each coordinate basis vector field $\partial_A$ of $\{\xi^A\}$ can be decomposed into
two parts on $D_4$: one part is tangent to $D_4$, denoted by $\widetilde{\partial}_A{}^a$,
and the other part is orthogonal to $D_4$, which is equal to $l^{-2}\xi_{A}\xi^{B}\partial_{B}$.
Similarly, each vertical coordinate basis vector field $\partial_A$ of $\{x^\mu,\xi^A\}$ can be
decomposed into two parts on ${\cal Q}$:
\begin{equation}\label{partialA}
\partial_{A}=\widetilde{\partial}_{A}+l^{-2}\xi_{A}\xi^{B}\partial_{B},
\end{equation}
where $\widetilde{\partial}_{A}$ is the pushforward of $\widetilde{\partial}_A{}^a$
by the map $\xi\rightarrow \sigma(x)\cdot\xi$, and is tangent to the fiber of ${\cal Q}$,
where $\xi\in D_4$. Suppose that $\theta^A$ are several vertical one-form fields on
$\{p\cdot\xi\,|\,p\in\pi^{-1}[\,{\cal U}\,],\xi\in D_4\}$,
where `vertical' means $\theta^A(X_\mu)=0$.
Assume that for any $x\in{\cal U}$, the pullbacks of $\theta^A$ by the map
$\xi\rightarrow \sigma(x)\cdot\xi$
are $d_a\xi^A$, then there should be \cite{Lu13}
\begin{equation}\label{theta}
\theta^{A}=d\xi^{A}+\Omega^{A}{}_{B\mu}\xi^{B}dx^{\mu},
\end{equation}
where $x^\mu$ and $\xi^A$ are coordinates of $\{x^\mu,\xi^A\}$.
Under the gauge transformation (\ref{gaugeT}), $\theta^A$ transform as
\begin{equation}
\theta'^A=[g(x)]^A{}_B\,\theta^B.
\end{equation}
A comparison of Eqs. (\ref{Daphi}) and (\ref{theta}) yields
\begin{equation}
(\phi^*\theta^A)_a=D_a\xi^A.
\end{equation}

Suppose that $\{E_\alpha{}^a\}$ is a local frame field on $D_4$, with $\{E^\alpha{}_a\}$ as its dual.
Then there exist functions $E_\alpha{}^A$, $E^\alpha{}_A$,
$E_A{}^\alpha$ and $E^A{}_\alpha$, such that $E_\alpha{}^a=(E_\alpha{}^A\partial_A)^a$,
$E^\alpha{}_a=E^\alpha{}_Ad_a\xi^A$, $\widetilde{\partial}_A{}^a=E_A{}^\alpha E_\alpha{}^a$
and $d_a\xi^A=E^A{}_\alpha E^\alpha{}_a$. Since $E_\alpha{}^a$ is tangent to $D_4$,
there should be
\begin{equation}\label{Exi}
E_\alpha{}^A\xi_A=0.
\end{equation}
Moreover, with the help of $E_\alpha{}^aE^\beta{}_a=\delta^\beta{}_\alpha$,
$E_\alpha{}^ad_a\xi^A=E_\alpha{}^A$, $E^\alpha{}_a\widetilde{\partial}_A{}^a=E_A{}^\alpha$
and $(d_a\xi^A)\widetilde{\partial}_B{}^a$ $=\delta^A{}_B-l^{-2}\xi^A\xi_B$,
it can be shown that
\begin{equation}\label{EE}
E_\alpha{}^AE^\beta{}_A=\delta^\beta{}_\alpha,\
E^A{}_\alpha E_B{}^\alpha=\delta^A{}_B-l^{-2}\xi^A\xi_B,
\end{equation}
\begin{equation}\label{EandE}
E^A{}_\alpha=E_\alpha{}^A,\
E_A{}^\alpha=E^\alpha{}_B(\delta^B{}_A-l^{-2}\xi^B\xi_A).
\end{equation}
Assume that $E_\alpha$ are several local vertical vector fields on ${\cal Q}$,
which are the pushforwards of $E_\alpha{}^a$ by the map $\xi\rightarrow\sigma(x)\cdot\xi$,
and $E^\alpha$ are several local vertical one-form fields on ${\cal Q}$,
whose pullbacks by the map $\xi\rightarrow\sigma(x)\cdot\xi$ are $E^\alpha{}_a$ for any $x\in{\cal U}$.
The local functions $E_\alpha{}^A$, $E^\alpha{}_A$, $E_A{}^\alpha$ and $E^A{}_\alpha$
on $D_4$ induce some local functions on ${\cal Q}$, which are denoted by the same symbols,
and defined by, for example,
\begin{equation}\label{EalphaA}
E_\alpha{}^A[\sigma(x)\cdot\xi]=E_\alpha{}^A(\xi).
\end{equation}
Obviously, these induced functions also satisfy Eqs.
(\ref{Exi}), (\ref{EE}) and (\ref{EandE}).
Furthermore, it can be verified that
\begin{equation}\label{Ealpha}
E_{\alpha}=E_{\alpha}{}^{A}\partial_{A},\
E^{\alpha}=E^{\alpha}{}_{A}\theta^{A},
\end{equation}
\begin{equation}\label{tpartialA}
\widetilde{\partial}_A=E_A{}^\alpha E_\alpha,\
\theta^A=E^A{}_\alpha E^\alpha,
\end{equation}
\begin{equation}
E^\alpha(E_\beta)=\delta^\alpha{}_\beta,\
\theta^A(\tilde{\partial}_B)=\delta^A{}_B-l^{-2}\xi^A\xi_B,
\end{equation}
\begin{equation}
E^\alpha(\tilde{\partial}_A)=E_A{}^\alpha,\
\theta^A(E_\alpha)=E_\alpha{}^A.
\end{equation}
Let $\{\overline{E}_A\}=\{X_\mu,E_\alpha\}$ and
$\{\overline{E}^A\}=\{dx^\mu,E^\alpha\}$. Then $\{\overline{E}_{\mathcal {A}}\}$ is a local
frame field on ${\cal Q}$,
with $\{\overline{E}^{\mathcal {A}}\}$ as its dual frame field.

\end{document}